\documentclass[twocolumn]{aa}
\usepackage{graphicx}
\usepackage{txfonts}
\usepackage{longtable}

\usepackage{natbib}

\newcommand{\yohkoh}{{\sl Yohkoh }}

\begin{document}
   \title{YOHKOH remnants: partially occulted flares in hard X-rays}

   \author{M. Tomczak}

   \offprints{M. Tomczak}

   \institute{Astronomical Institute, University of Wroc{\l }aw,
              ul. Kopernika 11, PL-51--622 Wroc{\l }aw, Poland,
              \email{tomczak@astro.uni.wroc.pl}
             }

   \date{Received ; accepted }

   \abstract
   {Modern solar X-ray imagers did not breakthrough the problem of
   detailed diagnostics of faint hard X-ray sources in the presence of stronger
   ones. This is the case of the impulsive phase of solar flares in which footpoint sources are usually
   stronger than loop-top ones.}
   {For this aim, flares being  partially occulted by the solar limb, are the best
   reservoir of our knowledge about hard X-ray loop-top sources. Recently,
   the survey of partially occulted flares observed by the RHESSI has been published (Krucker
   \& Lin 2008). The extensive \yohkoh database still awaits such activities. This
   work is an attempt to fill this gap.}
   {Among from 1286 flares in the \yohkoh Hard X-ray Telescope Flare Catalogue (Sato
   et al. 2006), for which the hard X-ray images had been enclosed, we identified 98
   events that occurred behind the solar limb. We investigated their hard X-ray spectra
   and spatial structure.}
   {We found that in most cases the hard X-ray spectrum of partially occulted flares
   consists of two components, non-thermal and thermal, which are co-spatial. The photon
   energy spectra of the partially occulted flares are systematically steeper than spectra
   of the non-occulted flares. Such a difference we explain as a consequence of intrinsically
   dissimilar conditions ruling in coronal parts of flares, in comparison with the footpoints
   which usually dominate the hard X-ray emission of disk flares. At least two reasons of the
   difference should be taken into consideration: (1) stronger contamination of hard X-rays with
   emission of thermal plasma, (2) different mechanism in which non-thermal electrons
   radiate their energy. For events unbiased with the thermal component the
   difference, ${\Delta}{\gamma} = \bar{{\gamma}}_{LT} - \bar{{\gamma}}_{FP}$, is equal to 1.5.}
   {A schematic picture, in which thin-target mechanism is responsible for hard X-ray emission of
   loop-top sources and thick-target mechanism -- for emission of footpoint sources, is modified by
   the presence of some coronal thick-target sources. At least a part of them suggests a magnetic trapping.
   Investigated flares do not respond the overall (global) magnetic configuration of
   the solar corona. For their characteristics conclusive is rather the
   local magnetic configuration in which they were developed.}

   \keywords{Sun: corona -- flares -- particle emission -- X-rays, gamma rays}

   \titlerunning{YOHKOH remnants: partially occulted flares in hard X-rays}

   \maketitle
%

\section{Introduction}
\label{intr}

It has been commonly accepted that solar flares are caused by
reconnection of magnetic field lines in the corona. In this process,
energy originally stored in the magnetic field is redistributed into
plasma heating, waves generation and particles acceleration.
However, details of reconnection process as well as general rules of
energy partition are subjects of extensive debate. Any progress
strongly depends on wise-organized observations and their
interpretation.

Hard X-ray observations offer a good insight into further evaluation
of particles accelerated in the reconnection process. Propagation of
particles operates under guidance of magnetic lines. Their bundles
converge at the entrance into the lower part of the solar
atmosphere, where density of ambient plasma increases steeply.
Electrons accelerated in the corona are stopped there emitting
intense hard X-radiation via electron-ion bremsstrahlung. This
mechanism, known as the thick-target model \citep{bro71}, works so
efficiently that footpoint hard X-ray sources usually strongly
dominate spatial distribution of flare emission in this energy
range.

It is difficult to observe the particles acceleration without
effects introduced by the propagation of particles because in the
less-dense corona hard X-radiation is emitted less efficiently in
the thin-target model \citep{bro71,lin74}. In many cases the coronal
hard X-ray sources can be seen as an effect of magnetic convergence
\citet{t+c07} or ultra-dense thick-target environment
\citep{kos94,v+b04}.

Coronal hard X-ray sources can be easily separated when a flare
occurs behind the solar limb but close enough for recording emission
of the higher part of magnetic structure. In case of a partially
occulted flare the solar disk works like a rough imager which stops
emission of usually brighter footpoint sources. Such a configuration
had been used routinely for investigation of coronal hard X-ray
sources before the hard X-ray imaging detectors began to operate
\citep{zir69,f+d71,hud78,hud82,kan82}.

First statistical attempt of partially occulted flares has been
performed using data from the UCSD experiment onboard {\sl OSO-7}
satellite \citep{r+d75,ken75}. In fifteen months of {\sl OSO-7}
operation, from among 601 X-ray bursts above
10$^3$\,photon\,cm$^{-2}$\,s$^{-1}$\,keV$^{-1}$ in energy channel
extending over 5.1-6.6 keV, 54 bursts were unaccompanied by
H$\alpha$ flares. As a cinema flare patrol was in progress during
those bursts, they probably occurred behind the solar limb.

From the group of partially occulted flares, \citet{ken75} has
chosen eight major soft X-ray events and has found that all had
significant hard X-ray emission in the 30-44 keV range.
\citet{r+d75} analyzed all available hard X-ray spectra of partially
occulted flares. They found that 25 from among 37 bursts, had a
non-thermal component. The average value of the spectral indices at
peak 20 keV flux for these 25 over-the-limb events was 4.6, whereas
for 59 center events ($0^{\circ} < \theta <60^{\circ}$) was 3.8.

The main conclusion of above mentioned papers from the pre-imaging
era was that hard X-ray emission was not concentrated close to the
solar surface but took place in extended regions in the corona.

Modern hard X-ray telescopes onboard \yohkoh (Hard X-ray Telescope)
and {\sl Reuven Ramaty High-Energy Solar Spectroscopic Imager
(RHESSI)} satellites have opened a new perspective in investigation
of coronal hard X-ray sources. Many important discoveries have been
done \citep{kru08a}, including the most famous -- the discovery of
the presence of above the loop-top sources in flares \citep{mas+94}.
However, in many cases a low dynamical range of a hard X-ray
telescope makes a qualitative analysis of images very problematic.
It considers the problem of detailed diagnostics of faint hard X-ray
sources in the presence of stronger ones. \citet{a+m97} have shown
that weak sources can be suppressed during the image reconstruction
and in consequence they can mimic the appearance of stronger
neighbors. As we see, this is the case of the impulsive phase of a
typical flare when we observe strong footpoint and weak loop-top
sources simultaneously.

To avoid such a complication the partially occulted flares were
chosen. From the \yohkoh database, one or a few examples of
behind-the-limb flares were selected many times so far.  This
specific configuration was usually used to perform joint diagnostics
on the basis of hard and soft X-ray images including full-Sun soft
X-ray spectra recorded by the Bragg Crystal Spectrometer. In this
way, soft X-ray bright loop-top kernels
\citep{kha95,mar96,ste96,m+d97,o+s97} as well as flare-associated
X-ray plasma ejections \citep{o+s97,tom04,tom05} were investigated.

\citet{m+m99} have done a comparison of hard and soft X-ray
characteristics for 28 partially occulted and 17 non-occulted limb
flares observed by \yohkoh between 1991 and 1996. Partially occulted
flares in most observational characteristics were found to be
indistinguishable from non-occulted ones. Exceptions were the hard
X-ray spectra averaged over the entire events which exhibit a higher
value of the power-law index $\gamma$ in the partially occulted
flares.

\citet{tom01} has investigated the \yohkoh X-ray images of 14
behind-the-limb flares. He reported a complex variability of hard
X-ray flux built by two separate components: several-minutes gradual
lower-energy backgrounds and quasi-periodic higher-energy impulses
lasting typically 5--30\,s. Impulses were too weak for imaging but
gradual components provides the easy identification in hard X-ray
images. The hard X-ray sources were co-spatial with soft X-ray
kernels (more frequent type A) as well as had no distinct
counterparts in soft X-rays (less frequent type B). The appearance
of a new gradual component in the hard X-ray light curve was always
associated with the presence of an additional hard X-ray source.

Observations of the partially occulted flares were recently analyzed
on the basis of {\sl RHESSI} data \citep{kru07a,l+g07,liu08},
together with {\sl Hinode} \citep{kru07b} or {\sl Solar-Terrestrial
Relations Observatory, STeReO} \citep{kru09} data. \citet{k+l08}
have prepared a survey of partially occulted flares observed by {\sl
RHESSI}. They found 55 such events between 2002 February and 2004
August. The existence of two different components of coronal hard
X-ray emission was revealed in 50 flares: (1) the thermal with a
gradual time profile, and (2) the non-thermal (power-law spectra
with indices $\gamma$ mostly between 4 and 7) showing faster time
variations. Both components were usually co-spatial within
${\sim}2{\times}10^3$\,km with only a few exceptions.

The extensive \yohkoh database still awaits preparing a survey of
partially occulted flares. This paper is an attempt to fill this
gap. For this aim, we used the Hard X-ray Telescope Flare Catalogue.
The development and structure of this catalogue are described in
Sect.\,\ref{cat}. The criteria of selection of flares enclosed in
the survey are given in Sect.\,\ref{crit}. In Sect.\,\ref{res} a
statistical approach of some characteristics describing partially
occulted flares is given and the results are compared to their
counterparts obtained for non-occulted flares from the Catalogue. In
Sect.\,\ref{dis} the results are discussed and compared to reports
of other authors. In Sect.\,\ref{inter} short descriptions of some
interesting groups of events in the Catalogue are given. Main
conclusions are summarized in Sect.\,\ref{concl}.

\begin{table*}
\caption[ ]{Development of the YOHKOH HXT Flare Catalogue}
\begin{flushleft}
\label{dev}
\begin{tabular}{ccccccccc}
 \hline
 Ver- & & Latest & & \multicolumn{5}{c}{Contents} \\
 \cline{5-9}
 sion & References & event & Records & HXT & HXT & {\sl GOES} & SXT & WBS \\
  & & & & (lc)$^{\rm a}$ & (i)$^{\rm a}$ & & & \\
  \hline
  1 & Kosugi et al. (1993) & 1992/12 & \hspace*{1.5mm}672 & + & -- & + & -- & -- \\
  2 & Kosugi et al. (1995) & 1994/12 & 1007 & + & -- & + & -- & -- \\
  3 & Sato et al. (1998) & 1998/08 & 1264 & + & + & + & -- & -- \\
  4 & Sato et al. (2003) & 2001/12 & 3112 & + & + & + & + & -- \\
  5 & Sato et al. (2006) & 2001/12 & 3112 & + & + & + & + & + \\
 \hline
\end{tabular}
\begin{list}{}{}
\item[$^{\rm a}$] lc -- light curves; i -- images
\end{list}
\end{flushleft}
\end{table*}

\begin{figure*}
\resizebox{\hsize}{!}{\includegraphics{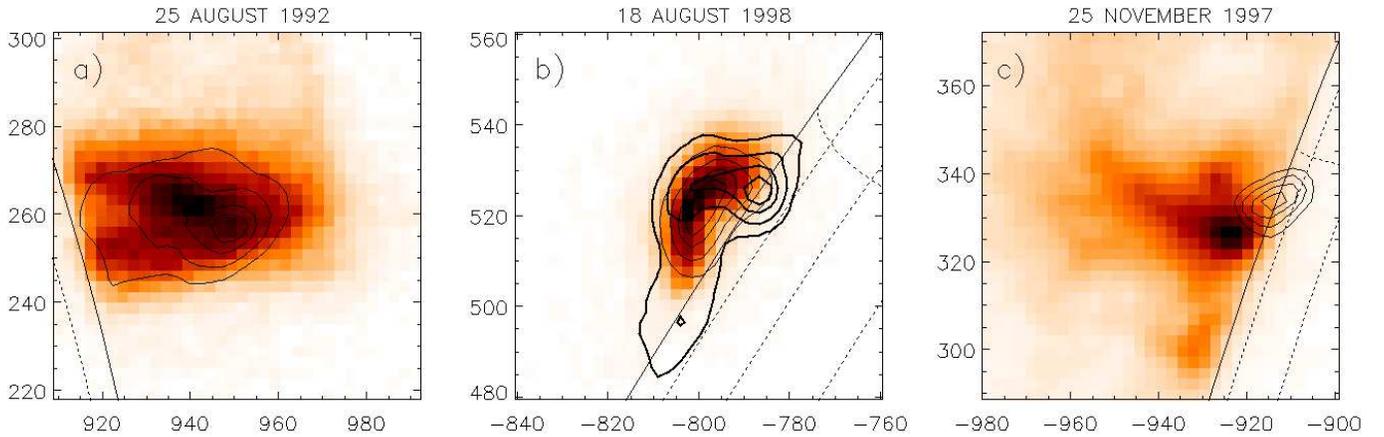}} \hfill
\caption{Three examples of events that were considered during the
preparation of the survey of partially occulted flares. Soft X-ray
images (halftones) are made by the SXT in the Be119 filter, thin
contours (10, 25, 50, and 75\% of $I_{max}$) represent hard X-ray
emission in energy band L (14-23 keV). Solid line shows solar limb
and dashed lines represent heliospheric coordinates on the solar
disk. {\bf a} Event qualified to the survey showing a typical
coronal source. {\bf b} Event excluded from the survey due to the
presence of footpoint sources seen in energy band M2 (33-53 keV) --
thick contours. {\bf c} Event qualified to the survey but excluded
from statistical considerations due to a possible footpoint emission
overshooting (estimated longitude $\theta$ =
E91$^{\circ}.6{\pm}0^{\circ}.6$).} \label{3ex}
\end{figure*}

\section{Development and structure of the YOHKOH HXT Flare Catalogue}
\label{cat}

The Japanese solar satellite \yohkoh operated in years 1991--2001
and during this time provided huge amount of excellent data. The
Hard X-ray Telescope, HXT, \citep{kos91} was a Fourier synthesis
imager observing the whole Sun. It consisted of 64 independent
subcollimators which measured spatially modulated intensities in
four energy bands (L: 14--23~keV, M1: 23--33~keV, M2: 33--53~keV,
and H: 53--93~keV). During the flare the intensities were
integrated, in each energy band, over 0.5~s. Some reconstruction
routines (Maximum Entropy Method, Pixons) that allow us to obtain
hard X--ray images with an angular resolution of up to 5 arcsec are
available.

Among from four scientific instruments onboard {\sl Yohkoh}, the
observations made by the HXT were organized in the most friendly way
since early years of the mission. The members of the HXT team have
prepared five versions of a catalogue collecting basic information
about flares observed by the telescope (see Table\,\ref{dev}). The
first two versions of the catalogue \citep{kos93,kos95} contain 672
and 1007 records, respectively. For some flares the 10-minutes hard
X-ray light curves in four energy bands are given. In the 3rd
version of the catalogue \citep{sat98}, containing 1264 records,
example hard X-ray images were added. The last two versions of the
catalogue contain the whole-mission database with 3112 records and
some supplementary data from other \yohkoh instruments. In the 4th
version \citep{sat03} the example soft X-ray images provided by the
Soft X-ray Telescope (SXT) were introduced. In the 5th version
\citep{sat06} light curves and spectra provided by the Wide Band
Spectrometer (WBS) were introduced.

In summary, the recent 5th version of the catalogue gives a short
description of 3090\footnote{Note that a total number of flares is
slightly lower than the total number of records because more than
one record is dedicated for some flares.} flares that produced at
least 3\,counts\,s$^{-1}$ per subcollimator (SC$^{-1}$) in channel
L. The description contains date, time, peak counts in four HXT and
four WBS energy bands, {\sl GOES} class, H$\alpha$ position, and
NOAA active region number. For 1286 flares example images in
particular energy bands, basically for a peak time period, are
enclosed. The criterion was a count rate roughly
10\,counts\,s$^{-1}$\,SC$^{-1}$ above the background. The HXT images
are overlaid on a SXT image when available. For the same 1286 flares
the four HXT light curves are presented. Moreover, the catalogue
contains WBS light curves of 2736 flares. Example WBS spectra of 369
flares up to 1\,MeV and spectra of 12 flares up to 10\,MeV are added
basically for a peak time period.

The last version of the catalogue still needs some improvements and
corrections. For example, for years 1997--2001 an identification of
{\sl GOES} class, H$\alpha$ position, and NOAA active region number
is made only for a minor part of events. The poor quality of some
HXT images suggests their probably wrong position, which is
supported by a disagreement with H$\alpha$ position of a flare.

\section{Criteria of selection} \label{crit}

In the first step, all hard X-ray images available in the Catalogue
were inspected. As a result, the preliminary list of 170 events,
located close to the solar limb, was established. In the next step,
each event from the preliminary list was carefully verified using
original files of data. For flares that occur behind the solar limb
the footpoint hard X-ray sources should not be seen. Therefore,
every flare from the list, for which a centroid of any hard X-ray
sources was located within the solar disk was rejected. If soft
X-ray images of the flare were available, decision of the rejection
was consulted with this kind of data. Namely, for flares that occur
behind the solar limb, there is a strong impression that the solar
limb sharply cuts the soft X-ray emission. Moreover, the
counterparts of footpoint hard X-ray sources are soft X-ray
brightenings \citep{hud94,tom97}. Thus, for events in our survey,
the soft X-ray brightenings should not be seen. If images of the
flare taken during the time interval chosen in the Catalogue did not
allow us to make the decision, we used the images for other time
interval. Three examples of flares considered during preparation of
the survey are presented in Fig.\,\ref{3ex}.

Finally, we have qualified 98 flares which observations strongly
suggest their situation behind the solar limb. The list of them is
presented in Table\,\ref{occlist}. How far behind the solar limb did
they occur? Acquaintance of heliographic longitude, $\theta$, is
crucial for a correct assignation of altitude above the photosphere,
$h$, one of the most important parameters for the later discussion.
However, before {\sl STeReO} satellites only limited estimations of

\begin{figure}
\resizebox{\hsize}{!}{\includegraphics{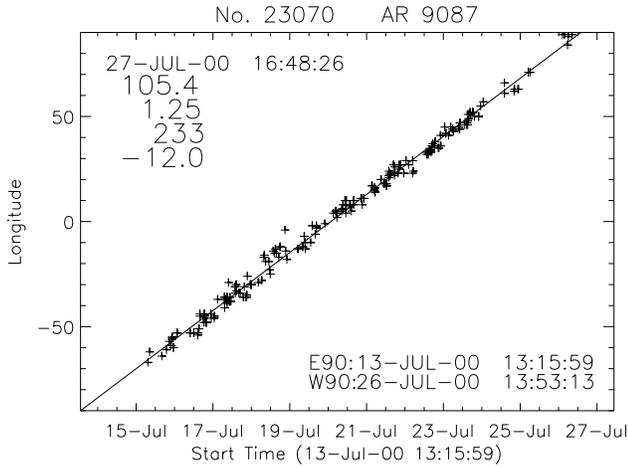}} \hfill
\caption{Example of the longitude extrapolation. Description on the
plot shows: time of event, value of $\theta$ and its uncertainty,
number of points, average value of the latitude (top, left), and
times of limb occurrences (bottom, right).} \label{pass}
\end{figure}the longitude were possible.

We estimate this parameter twofold. First, by using beyond-the-limb
extrapolation of a straight line fitted to the plot of longitude
versus time for all H$\alpha$ flares observed in the proper active
region during its passage across the solar disk \citep{r+d75}. For
this aim, observations collected in the {\sl Solar-Geophysical Data}
(SGD) were used. An example of the longitude extrapolation for the
flare No.\,23\,070 in the Catalogue from the NOAA AR 9087 is given
in Fig.\,\ref{pass}. Second, by calculating a position of the flare
on the basis of a time of the central meridian passage (CMP) taken
from the SGD for the proper active region. The formula describing
the solar differentional rotation given by \citet{n+n51} for large
recurrent sunspots, and corrected by {\citet{war66}, was used:
\begin{equation}
{\theta} = 13.45 - 3.0\,
{\sin}^2{\phi}\,\quad\mbox{deg\,day$^{-1}$,}\quad
\end{equation}
where $\phi$ is the region latitude.

Both methods of the estimation of $\theta$ work very well only under
specific conditions. The first one shows the position of a weighted
centroid of positions in which flares occurred in the active region
during its passage across the solar disk. The second one shows
basically the position of the central part of an active region.
Thus, if the investigated behind-the-limb flare occurred in the
alternative place of the active region, each method may be charged
with some additional (systematic) error. That is why sometimes the
obtained values of $\theta$ were below 90$^{\circ}$ though the
available X-ray images strongly suggested the position beyond the
solar limb.

The values of $\theta$ given in column (6) of Table\,\ref{occlist}
were obtained basically by the first method, except for the case
when the $\theta$-$t$ plot consisted of only a few points or the
second method provided a more realistic solution. Uncertainties
given in column (7) of Table\,\ref{occlist} are a measure of the
scatter of source data taken from the SGD. The results are
summarized in Fig.\,\ref{theta-phi}, where the latitude is taken
directly from the hard X-ray image. The obtained values of the
longitude are between slightly below 90$^{\circ}$ (a consequence of
systematic errors) to 26$^{\circ}$.9$\pm$1$^{\circ}$.8 beyond the
solar limb (event No.\,26\,060). For the event No.\,15\,640 only we
could not estimate the longitude.

At first glance, it seems that in our survey some partially occulted
flares are missed. We identified only 98 events out of 1286 i.e. 7.6
percent of the catalogue supplies. Assuming uniform distribution in
the longitude of all flares in the Catalogue one should expect that
4/22 i.e. 18.2\% of the flares have occurred behind the solar limb
between 0$^{\circ}$ and 20$^{\circ}$. We should remember, however,
that the further position of a flare beyond the solar limb, the
larger is a occultation height. For example, for 10$^{\circ}$ hard
X-ray emission should occur at least about
$\sim$1.1$\times$10$^4$\,km above the photosphere to be seen and for
20$^{\circ}$ the minimal height increases to about
$\sim$4.6$\times$10$^4$\,km. This effect seriously decreases the
population of partially occulted flares.

\begin{figure}
\resizebox{\hsize}{!}{\includegraphics{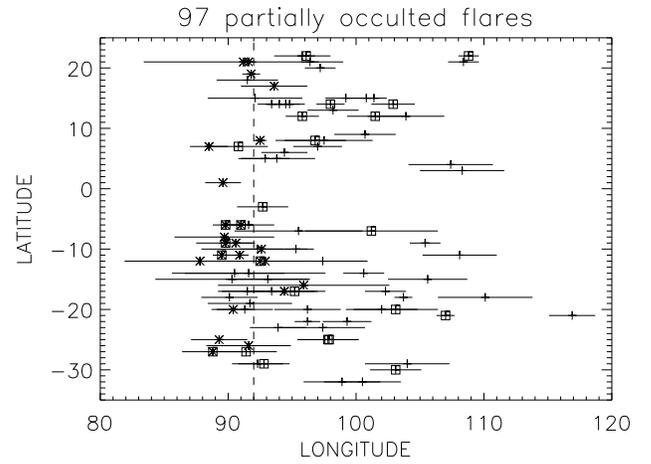}} \hfill
\caption{Plot of heliographic coordinates (longitude versus
latitude) for 97 flares from our survey. Events observed outside the
maximum of hard X-rays are marked with boxes, events with a
footpoint emission overshooting are marked with stars. The
termination longitude for the overshooting (92$^{\circ}$) is shown.}
\label{theta-phi}
\end{figure}

\begin{figure}
\resizebox{\hsize}{!}{\includegraphics{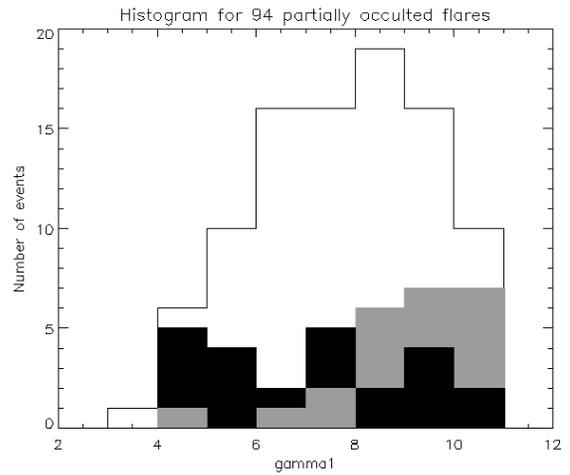}} \hfill
\caption{Histogram of values of power-law index obtained for 94
partially occulted flares in energy range 14-33\,keV. Black bins
represent 24 events observed outside the maximum of hard X-rays,
gray bins represent 24 events with the footpoint emission
overshooting.} \label{hist94}
\end{figure}

We tried to estimate the corrected population of behind-the-limb
flares by including frequency of their occurrence in dependence on
the altitude above the photosphere. We adopted the formula
\begin{equation}
\frac{dI}{dh} \sim \exp{(-\frac{h}{h_0})}
\end{equation}
where the scale-height $h_0 = (11 \pm 3) \times 10^3$\,km was
pointed out by \citet{c+a73} for soft X-rays. The similar
scale-height was concluded from the investigation of the peak time
differences between soft X-ray and hard X-ray emissions of 859
flares \citep{l+g06}. The obtained value (7.7$\pm$1.3)\% agrees
almost perfectly with the actual population (7.6\%) in the
Catalogue. We conclude that our survey is almost complete.

\section{Results}
 \label{res}

In our survey we characterize each event by its appearance during a
single time interval established in the HXT Flare Catalogue. This
time interval was chosen to contain the strongest hard X-ray signal.
Its duration depended on a number of counts and imaging
requirements, thus, sometimes it was only a few seconds around a
single burst, sometimes -- a lot of minutes with many bursts (see
columns (3) and (14) in Table\,\ref{occlist}).

\subsection{Power-law spectra}
 \label{pl}

Within this time interval the total signal above the background from
the whole image was accumulated in each energy band, which means
that our results are not charged with disadvantages of imaging
spectroscopy. We calculated flux ratios between consecutive energy
bands: M1/L, M2/M1, and H/M2, so called hardness ratios. From
hardness ratios the average power-law indices in some energy ranges
of the hard X-ray spectrum were obtained: M1/L $\rightarrow
{\gamma}1$, between 14 and 33 keV, M2/M1 $\rightarrow {\gamma}2$,
between 23 and 53 keV, and H/M2 $\rightarrow {\gamma}3$, between 33
and 93 keV.

Results are given in columns (11) - (13) of Table\,\ref{occlist}.
Many events showed a signal above the background only in low-energy
bands, L and M1, thus for them there are no values of ${\gamma}2$
and ${\gamma}3$ in the table. For three flares even the value of
${\gamma}1$ was not obtained: for two of them (Nos. 09\,650 and
24\,840) source data files were corrupted, for event No.\,22\,570
the hard X-ray flux was recorded only in the channel L.
Uncertainties of $\gamma$ indices, estimated according to number of
counts using the Poisson statistics, were typically 0.1--0.2.

\begin{figure}
\resizebox{\hsize}{!}{\includegraphics{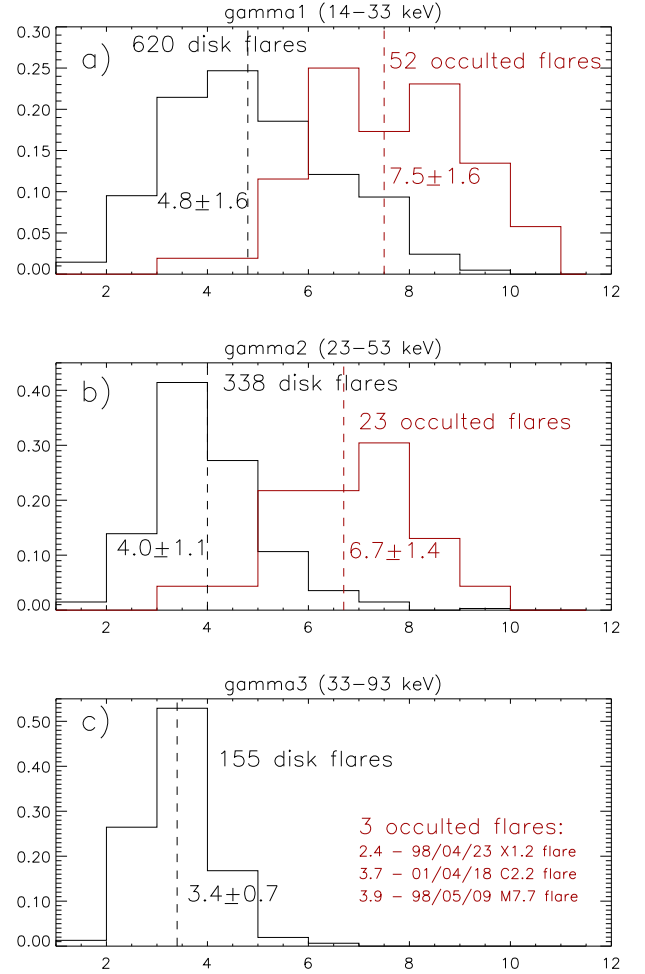}} \hfill
\caption{Normalized histograms of values of the power-law index
fitted in different ranges of the hard X-ray energy photon spectrum.
Separate histograms for partially occulted and disk flares are
presented. Energy ranges, number of events, average values and
standard deviations are written.} \label{histg}
\end{figure}

In Fig.\,\ref{theta-phi} two kinds of flares not included in further
statistical considerations are marked. The events marked with boxes
were observed outside the maximum of hard X-rays, therefore values
of $\gamma$ indices obtained for them are unrepresentative for the
phase of the highest energy available for other events. For
uniformity of results, these flares are omitted in the later
analysis.

The events marked in Fig.\,\ref{theta-phi} with stars occurred
probably too close to the solar limb and hard X-ray emission of
their footpoints is not eliminated completely. Indeed, for these
events a small hard X-ray source is seen centered close to the solar
limb (e.g. Fig.\,\ref{3ex}c). Its size is much smaller than a size
of a typical coronal source. We found that such sources are seen
basically for the longitudes below 92$^{\circ}$. Our aim is to study
characteristics of coronal hard X-ray sources, therefore these
flares also are omitted in later considerations.

A histogram of values of the ${\gamma}1$ index obtained for 94
events from our survey is presented in Fig.\,\ref{hist94}. Moreover,
two separate histograms: for events without the hard X-ray maximum
(boxes in Fig.\,\ref{theta-phi}) and for events with the footpoint
emission overshooting (stars in Fig.\,\ref{theta-phi}) are
overwritten. According to our expectations the first histogram,
containing less energetic events, is shifted towards higher values,
while the second one, including more energetic events, is shifted
towards lower values. We conclude that due to applied strict
criteria of the selection, the finally qualified population is more
uniform.

\begin{figure}
\resizebox{\hsize}{!}{\includegraphics{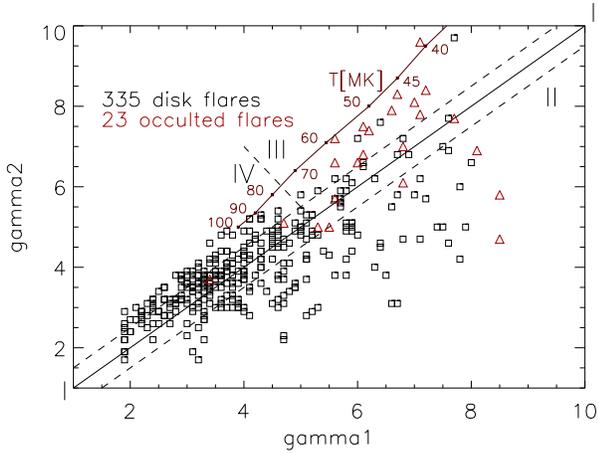}} \hfill
\caption{Comparison between values of the power-law index fitted in
two energy ranges: 14-33\,keV (${\gamma}1$) and 23-53\,keV
(${\gamma}2$). Partially occulted flares are marked with triangles,
non-occulted flares are marked with boxes. In dependence on
mechanisms responsible for a hard X-ray spectrum formation the
points are situated in one of four areas described by digits between
I and IV. For additional explanations -- see text. Statistics of
occurrence is presented in Table\,\ref{st:g1g2}. The solid narrow
line refers to emission of purely thermal plasma with temperatures
between 38 and 100 MK.} \label{g1g2}
\end{figure}

\begin{table}
\caption[ ]{Statistics of different types of spectra in 14-53\,keV
energy range}
\begin{flushleft}
\label{st:g1g2}
\begin{tabular}{cccc}
 \hline
 Area & Type of & Partially & Non-occulted \\
 in Fig.\,\ref{g1g2} & spectrum & occulted & \\
  \hline
  I & non-thermal & 30.4\% & 55.2\% \\
   & (${\gamma}1 \sim {\gamma}2$) & (7/23) & (185/335) \\
  II & thermal + non-thermal & 17.4\% & 24.2\% \\
   & (${\gamma}1 > {\gamma}2$) & (4/23) & (81/335) \\
  III & quasi-thermal & 52.2\% & 1.8\% \\
   & (${\gamma}1 < {\gamma}2$) & (12/23) & (6/335) \\
  IV & albedo & 0\% & 18.8\% \\
   & (${\gamma}1 < {\gamma}2$) & (0/23) & (63/335) \\
 \hline
\end{tabular}
\end{flushleft}
\end{table}

The final histogram is presented in Fig.\,\ref{histg}a. It contains
values for 52 partially occulted flares which we have left after the
rejection of two groups of events described in the previous
paragraph. The histogram of values of the ${\gamma}1$ index obtained
for 620 non-occulted flares found in the Catalogue is given in the
same panel. For easy comparison both histograms are normalized. As
we can see, central parts of histograms, containing about two third
of events, are shifted one from another. The average value for
partially occulted flares is 7.5, while the average value for
non-occulted flares is 4.8. Standard deviations for both histograms
are 1.6.

In Fig.\,\ref{histg}b we compared normalized histograms of the
${\gamma}2$ index obtained for partially occulted and non-occulted
flares, respectively. Because for a part of flares signal above the
background in energy band M2 was imperceptible, the total number of
events has decreased to 23 and 338, respectively. Again, the central
parts of histograms, containing above 70\% of events, are shifted
one from another.  The average value for partially occulted flares
is 6.7, while the average value for non-occulted flares is 4.0.
Standard deviations are 1.4 and 1.1, respectively.

Calculation of the ${\gamma}3$ index needs a distinct signal above
the background in energy band H. This limits once more a number of
events to 155 non-occulted and only 3 partially occulted flares.
Therefore, in Fig.\,\ref{histg}c we present only the normalized
histogram for non-occulted flares. The average value is 3.4 and the
standard deviation is 0.7. Three partially occulted flares emitting
signal strong enough in energy band H show values of the ${\gamma}3$
index similar to typical values obtained for disk flares. They are:
event No.\,11\,650 of 1998 April 23 \citep{sat01,tom04}, event
No.\,11\,950 of 1998 May 9 \citep{tom08}, and event No.\,26\,060 of
2001 April 18 \citep{hud01}. We would like to stress that all these
events show very spectacular X-ray plasma ejections.

\subsection{Deviations from power-law spectra}
 \label{therm}

Histograms of values of the power-law index in Fig.\,\ref{histg}
show a shift towards lower values with an increase of photon
energies for which the power-law was fitted. This shift is seen for
partially occulted as well as for non-occulted flares. Such
appearance strongly suggest a contamination of hard X-rays with
emission of thermal plasma which is expected to be decreasing with
the photon energy.

The HXT provides poor spectral resolution (only four broad energy
bands) which excludes a reasonable fitting of a non-thermal
power-law component together with a thermal one. Therefore, another
method to distinguish between non-thermal and thermal components
should be proposed. In this paper we calculated the average
power-law index ${\gamma}1$, ${\gamma}2$, and ${\gamma}3$ for energy
ranges: 14-33, 23-53, and 33-93\,keV, respectively. If an
investigated photon energy spectrum could be described with a single
power-law formula then the obtained values of ${\gamma}1$,
${\gamma}2$, and ${\gamma}3$ should be similar within observational
uncertainties. Some larger differences between these values strongly
suggest a deviation from the single power-law caused, for example,
by the presence of the thermal component.

\begin{figure}
\resizebox{\hsize}{!}{\includegraphics{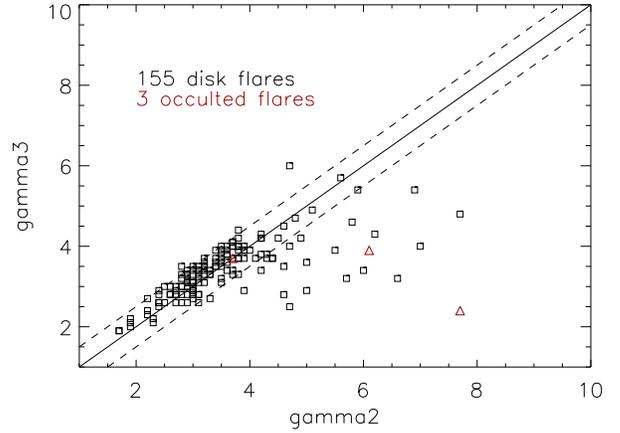}} \hfill
\caption{Comparison between values of the power-law index fitted in
two energy ranges: 23-53\,keV (${\gamma}2$) and 33-93\,keV
(${\gamma}3$). Partially occulted flares are marked with triangles,
non occulted flares are marked with boxes. Statistics of occurrence
is presented in Table\,\ref{st:g2g3}. For other explanations -- see
text.} \label{g2g3}
\end{figure}

\begin{table}
\caption[ ]{Statistics of different types of spectra in 23-93\,keV
energy range}
\begin{flushleft}
\label{st:g2g3}
\begin{tabular}{ccccc}
 \hline
 Area & Type of & Partially &  & Non- \\
 in Fig.\,\ref{g2g3} & spectrum & occulted & & occulted \\
  \hline
  I & non-thermal & & 82.9\% & \\
   & (${\gamma}2 \sim {\gamma}3$) & (1/3) & & (130/155) \\
  II & thermal + non- & & 15.2\% & \\
   & thermal (${\gamma}2 > {\gamma}3$) & (2/3) & & (22/155) \\
  III & quasi-thermal & & 0\% & \\
   & (${\gamma}2 < {\gamma}3$) & (0/3) & & (0/155) \\
  IV & albedo & & 1.9\% & \\
   & (${\gamma}2 < {\gamma}3$) & (0/3) & & (3/155) \\
 \hline
\end{tabular}
\end{flushleft}
\end{table}

In Fig.\,\ref{g1g2} we present the comparison between ${\gamma}1$
and ${\gamma}2$ for flares from histograms in Fig.\,\ref{histg}a-b.
For events situated within the central corridor between two straight
dashed lines (area I), differences between ${\gamma}1$ and
${\gamma}2$ are below expected uncertainties: $|{\gamma}1 -
{\gamma}2| \le 0.5$. It means that in energy range 14-53\,keV their
spectra can be described by the single power-law. We consider them
as purely non-thermal.

Area II below the central corridor is occupied by flares for which
$({\gamma}1 - {\gamma}2) > 0.5$. It means that their spectra are
steeper in lower and flatter in higher energies. We explain this
shape as the consequence of the thermal component presence in energy
band L.

The area above the central corridor is occupied by flares for which
$({\gamma}2 - {\gamma}1) > 0.5$. It means that their spectra are
flatter in lower and steeper in higher energies. This shape suggests
a thermal nature. However, a purely thermal events should be located
in the figure along the solid narrow line above the corridor. As we
can see, only a few events is situated close to this line. The
majority of events located in area III is rather quasi-thermal i.e.
the thermal component is dominating but mixed with non-thermal
bursts as was reported by \citet{tom01}.

Events situated in area IV are too energetic to explain their
spectra by means of thermal emission. \citet{z+h03} proposed that
the value of ${\gamma}1$ lower than ${\gamma}2$ can be caused by the
Compton backscattering. This effect known as the photospheric albedo
moves a part of photons from energy band L to higher energies.
\citet{kas07} have shown that the photospheric albedo depends on the
heliocentric distance. We checked it and found that longitudes of 38
events out of 50 from area IV, for which H$\alpha$ coordinates are
given in the Catalogue, are below 45$^{\circ}$. We consider this
statement as a supporting argument that events situated in area IV
has non-thermal spectra modified by the photospheric albedo.

\begin{figure}
\resizebox{\hsize}{!}{\includegraphics{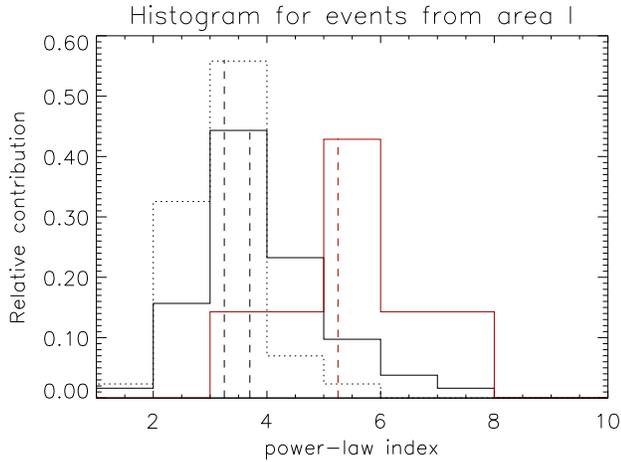}} \hfill
\caption{Normalized histograms of values of the power-law index for
flares situated in area I of Figs.\,\ref{g1g2} and \ref{g2g3} which
are believed to be free of disturbances introduced by the thermal
component and the photospheric albedo. Solid lines represent the
data for energy range 14-53 keV for partially occulted (red) and
non-occulted (black) flares separately. The dotted line represents
the data for energy range 23-93 keV for non-occulted flares. Medians
of histograms are marked with dashed lines.} \label{area1}
\end{figure}

Statistics of the occurrence of flares in different areas in
Fig.\,\ref{g1g2} for partially occulted and non-occulted flares
separately is given in Table\,\ref{st:g1g2}. The results can be
summarized as follows. During the hard X-ray maximum in energy band
L (14-23 keV) the non-thermal component dominates in about 74\% of
non-occulted flares and only in about 30\% of partially occulted
flares. On the other hand, the thermal component dominates in above
50\% of partially occulted flares and only in about 2\% of
non-occulted flares.

In Fig.\,\ref{g2g3} we present the comparison between values of
${\gamma}2$ and ${\gamma}3$ for flares from histograms in
Fig.\,\ref{histg}b-c. The shift of the investigation towards higher
energies reduces the heterogeneity of hard X-ray spectra seen in
Fig.\,\ref{g1g2}. The points are concentrated in areas I and II. It
means that in energy range 23-93 keV only pure non-thermal spectra
or thermal and non-thermal mixtures frequently occur. The statistics
is presented in Table\,\ref{st:g2g3}. Only total results are given
because there is a few partially occulted events and they show
values similar to non-occulted flares. As we can see, about 85\% of
flares during the hard X-ray maximum in energy band M1 (23-33 keV)
show purely non-thermal spectra, but the thermal component is still
present in about 15\% of events.

Let us check how much the deviations from power-law spectra have
disturbed the obtained values of power-law indices. The histograms
of values of the power-law index for flares situated in area I of
Figs.\,\ref{g1g2} and \ref{g2g3} are presented in Fig.\,\ref{area1}.
We expect that these events are free of disturbances introduced by
the thermal component and the photospheric albedo. The histogram of
185 non-occulted events from Fig.\,\ref{g1g2} resembles the
histogram shown in Fig.\,\ref{histg}b. Their medians are 3.7$\pm$1.1
and 4.0$\pm$1.1, respectively. This small difference is due to low
amounts of thermally dominated flares in case of non-occulted events
(see Table\,\ref{st:g1g2}).

Partially occulted flares is another case. Due to a frequent
occurrence of thermally dominated events (Table\,\ref{st:g1g2}) the
histogram shown in Fig.\,\ref{area1} is distinctly shifted towards
lower values in comparison with the histogram presented in
Fig.\,\ref{histg}b. The medians are 5.2$\pm$1.4 and 6.7$\pm$1.4,
respectively. In conclusion, after removing the contamination of
hard X-rays with emission of thermal plasma, partially occulted
flares in our survey still show higher values of the power-law index
than non-occulted flares. However, the difference between medians of
histograms in Figs.\,\ref{histg}b and \ref{area1} has decreased from
2.7 to 1.5.

Is it possible to eliminate in this way the contribution of thermal
emission completely? The comparison between histograms for
non-occulted flares in Fig.\,\ref{area1} for two energy ranges of
investigated spectra suggests that the interpretation should be done
with caution. The histogram obtained for higher energies is slightly
shifted towards lower values of the power-law index. Its median is
3.2$\pm$0.7 i.e. 0.5 lower than the median of the second histogram.
This can be interpreted as the cause of an uncompensate thermal
contribution, however, one cannot exclude another explanation,
namely the broken power-law with the break energy around 50 keV
(i.e. between energy bands M2 and H) and a flatter slope above this
value.

\subsection{Situation of coronal hard X-ray sources}
 \label{sit}

\begin{figure}
\resizebox{\hsize}{!}{\includegraphics{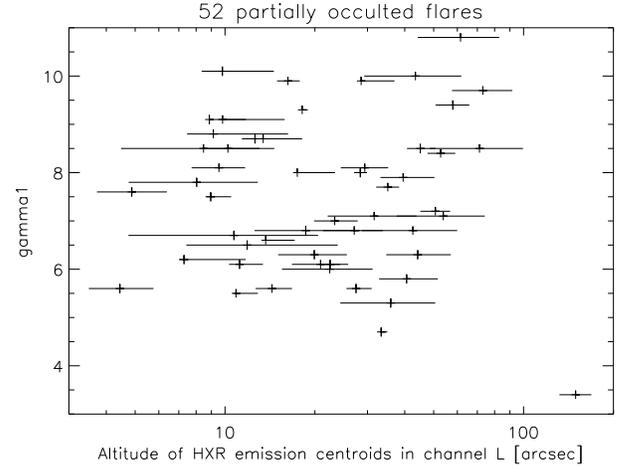}} \hfill
\caption{The power-law index ${\gamma}1$ measured between 14-33\,keV
against the altitude of the centroid of hard X-ray emission imaged
in energy band L for 52 partially occulted flares.} \label{h-g}
\end{figure}

In this subsection we use imaging ability of the HXT for
investigating the spatial structure of coronal sources. In first
step we have checked the dependence of observed hard X-ray spectra
on the altitude of coronal sources above the solar surface.

In Fig.\,\ref{h-g} we compared the power-law index ${\gamma}1$ to
the altitude of the centroid of hard X-ray emission imaged in energy
band L for 52 partially occulted flares. In case of multiple sources
the situation of the brightest one was taken into account. The
altitude was pointed out as a sum of two components: a height above
the solar limb $h_1$ (column 9 in Table\,\ref{occlist}) and a
occultation height $h_2$. The latter one was calculated on the basis
of heliospheric coordinates (column 6 in Table\,\ref{occlist}) as
follows:
\begin{equation}
h_2 = [{\sec}{\theta} - 1] {\sec}{\psi} R_{\sun},
\end{equation}
where $\theta$ is the longitude behind the solar limb and $\psi$ is
the latitude. The main source of the altitude error is the
uncertainty of the longitude (column 7 in Table\,\ref{occlist}). As
we can see, no correlation is observed -- flares showing hard X-ray
spectra having different shapes occur at each altitude. We found a
lack of correlation between the altitude and the power-law index
also for higher energy ranges.

How are the non-thermal and thermal components in investigated
flares spatially situated? To find out an answer, we compared the
situation of coronal hard X-ray sources to the situation of bright
loop-top kernels seen in soft X-rays. In Fig.\,\ref{shift} we
compared a shift between centroids of emission in these two ranges
of X-rays to the altitude of flares. We considered images of 41
partially occulted flares for which the hard X-ray image in energy
band L and the soft X-ray image taken in the Be119 filter were
obtained simultaneously. For a better clarity error bars are
omitted. The typical error of the shift was estimated to be about
2\,arcsec. The altitude error is the same as in Fig.\,\ref{h-g}.

\begin{figure}
\resizebox{\hsize}{!}{\includegraphics{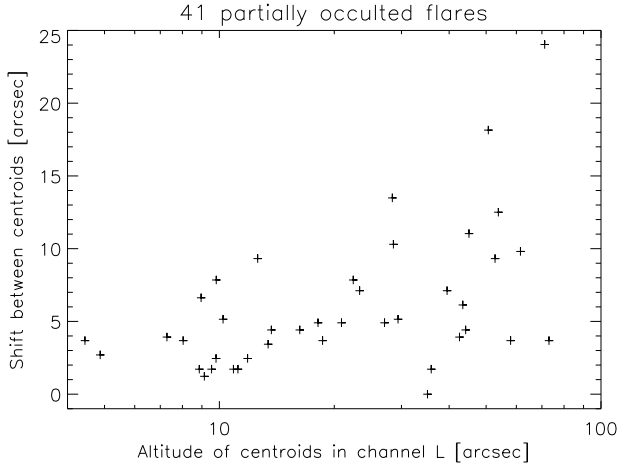}} \hfill
\caption{The shift between centroids of hard (HXT, energy band L)
and soft (SXT, Be119 filter) X-ray emission against the altitude of
the centroid of hard X-ray emission for 41 partially occulted
flares.} \label{shift}
\end{figure}

For 26 events (63.4\%) the spatial shift between the non-thermal and
thermal components is not higher than 5\,arcsec. The maximal values
of the shift seem to be increasing with the altitude. However, this
appearance can be caused by an increase with the altitude of
magnetic features sizes, e.g. soft X-ray loop-top kernels
\citep{p+k07}. To check this we have normalized the values presented
in Fig.\,\ref{shift} by dividing the shift by a characteristic size
of a hard X-ray source.

The range of values of the normalized shift seen in
Fig.\,\ref{nshift} was calculated as follows. We approximated the
hard X-ray sources at the level of 0.5$I_{max}$ as ellipses and
divided the shifts from Fig.\,\ref{shift} by the large and small
semi-axis of the ellipse. It establishes a lower and higher boundary
of the normalized shift, respectively. As we can see, the normalized
shift does not depend on the altitude and for the majority of events
(36 from 41) does not exceed the value 0.5. The actual co-spatiality
of hard X-ray and soft X-ray emission for some investigated events
is even better because small semi-axes (high boundaries of
normalized shifts) were shortened due to an occultation of the solar
limb.

\begin{figure}
\resizebox{\hsize}{!}{\includegraphics{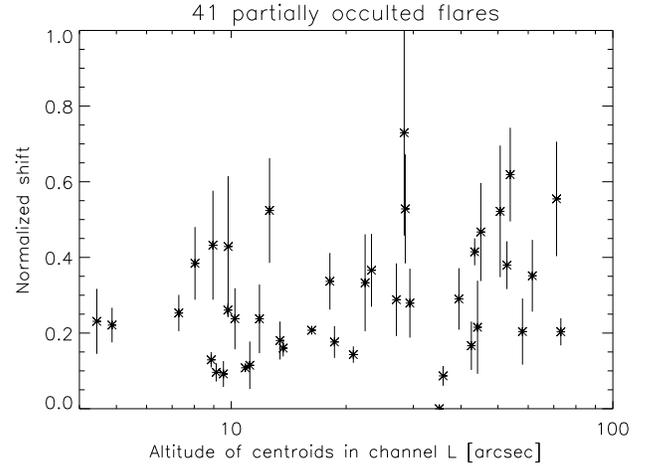}} \hfill
\caption{Shifts from Fig.\,\ref{shift} normalized by dividing with
characteristic sizes of hard X-ray sources in energy band L against
the altitude of flares for 41 partially occulted flares. For further
explanations -- see text.} \label{nshift}
\end{figure}

Figs.\,\ref{shift} and \ref{nshift} do not distinguish which kind of
emission, soft or hard X-ray, is situated higher. On the other hand,
this appearance is crucial to fit to some theoretical models.
Therefore, in Fig.\,\ref{radsep} we present the radial separation
between centroids of hard X-ray and soft X-ray emission for 41
partially occulted flares from Fig.\,\ref{shift}. Almost a half of
these events (21 from 41) show the same altitude of hard X-ay and
soft X-ray emission, within error bars. The second half show the
differences between them up to 14 arcsec with a supremacy of events
in which hard X-ray emission is located higher than soft X-ray
emission (15 from remaining 20). For 5 events soft X-ray emission
was situated higher than hard X-ray emission.

In Figs.\,\ref{h-g}-\ref{radsep} we used images taken in energy band
L as a representation of hard X-ray emission of investigated flares.
One can doubt in conclusions formulated on the basis of observations
in this energy band due to a strong contamination with emission of
thermal plasma. Perhaps, Figs.\,\ref{nshift}-\ref{radsep} suggest
co-spatiality of emissions in hard and soft X-rays as we observe
actually the same thermal plasma? To verify this doubt we compare
the situation of centroids of emission in energy band L to those
imaged in higher energies. Unfortunately, a smaller number of events
could be checked in this way, because for energy bands M1 and M2
were available only 16 and 5 images, respectively. We found that
only for two events, Nos.\,11\,950 and 25\,870, the radial
separation between emission in energy band L and in higher energies
is greater than 4\,arcsec. The remaining 14 events are almost
perfectly co-spatial.

\begin{figure}
\resizebox{\hsize}{!}{\includegraphics{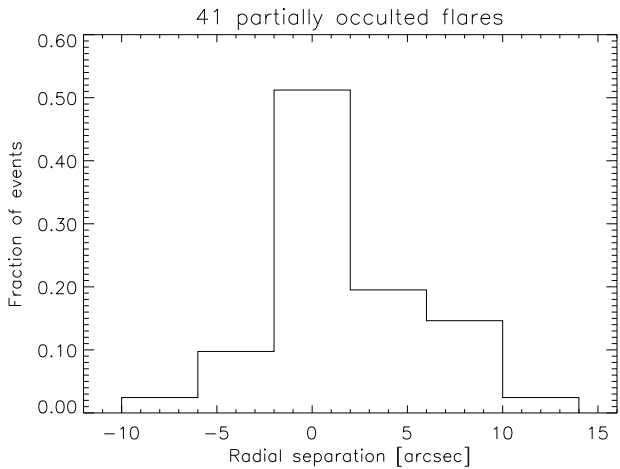}} \hfill
\caption{The radial separation between centroids of hard X-ray and
soft X-ray emission for 41 partially occulted flares from
Fig.\,\ref{shift}.} \label{radsep}
\end{figure}

\section{Discussion and interpretation of results}
 \label{dis}

Table\,\ref{tab:comp} collects the results of surveys in which hard
X-ray spectra of loop-top sources were investigated and compared to
spectra of footpoint sources. A large number of solar flares
included in these surveys allow us to expect that the results are
statistically important.

\begin{table*}
\caption[ ]{Surveys comparing power-law spectrum indices of hard
X-ray loop-top and footpoint sources}
\begin{flushleft}
\label{tab:comp}
\begin{tabular}{cccccccccccccccc}
 \hline
 & Time & Energy
 & \multicolumn{12}{c}{Power-law index ${\gamma}^{\rm a}$} & \\
 \cline{4-15} Satellite & period & range &
  \multicolumn{6}{c}{loop-top} & \multicolumn{6}{c}{footpoint} & References$^{\rm b}$ \\
 \cline{4-15} & [YY/MM] & [keV] & N & Mn & Md & SD & K & S & N & Mn & Md & SD & K &
 S & \\
  \hline
{\sl OSO-7} & 71/10-72/12 &  & 25 & 4.8 & 4.8 & 1.0 & -1.0 & \hspace*{1mm}0.1 & \hspace*{2mm}59 & 3.9 & 3.8 & 1.1 & -0.6 & 0.6 & (1) \\
\yohkoh & 91/10-96/05 & 14-33 & 25 & 8.2 & 8.1 & 1.5 & -1.1 & -0.1 & \hspace*{2mm}16 & 7.3 & 7.7 & 1.3 & -0.9 & \hspace*{-1mm}-0.4 & (2) \\
\yohkoh & 91/10-98/08 & 14-53 & \hspace*{1mm}12$^{\rm c}$ & 6.6 & 6.2 & 1.5 & \hspace*{1mm}0.2 & \hspace*{1mm}0.4 & \hspace*{2.5mm}18$^{\rm c}$ & 5.2 & 4.9 & 1.6 & -1.0 & 0.4 & (3) \\
\yohkoh & 91/10-98/08 & 14-93 & \hspace*{2.5mm}5$^{\rm c}$ & 5.5 & 5.3 & 1.3 & -1.5 & \hspace*{1mm}0.2 & \hspace*{2.5mm}13$^{\rm c}$ & 4.3 & 4.1 & 1.0 & -1.0 & 0.7 & (3) \\
\yohkoh & 91/10-01/12 & 14-33 & 52 & 7.5 & 7.6 & 1.6 & -0.5 & -0.1 & 620 & 4.8 & 4.6 & 1.6 & -0.6 & 0.4 & (4) \\
\yohkoh & 91/10-01/12 & 23-53 & 23 & 6.7 & 6.9 & 1.4 & -0.7 & -0.2 & 338 & 4.0 & 3.8 & 1.1 & \hspace*{1mm}2.1 & 1.0 & (4) \\
\yohkoh & 91/10-01/12 & 33-93 & \hspace*{1.5mm}3 & -- & -- & -- & -- & -- & 155 & 3.4 & 3.4 & 0.7 & \hspace*{1mm}1.2 & 0.6 & (4) \\
\yohkoh & 91/10-01/12 & 14-53 & \hspace*{2.5mm}7$^{\rm d}$ & 5.6 & 5.2 & 1.4 & -1.3 & \hspace*{1mm}0.2 & \hspace*{1mm}185$^{\rm d}$ & 3.9 & 3.7 & 1.1 & \hspace*{1mm}0.8 & 0.8 & (4) \\
\yohkoh & 91/10-01/12 & 23-93 & \hspace*{2.5mm}1$^{\rm d}$ & -- & -- & -- & -- & -- & \hspace*{1mm}130$^{\rm d}$ & 3.3 & 3.2 & 0.7 & \hspace*{1mm}1.4 & 0.6 & (4) \\
 {\sl RHESSI} & 02/02-04/08 & & 50 & 5.4 & 5.5 & 1.2 & -0.8 & \hspace*{1mm}0.3 & \hspace*{3mm}0 & -- & -- & -- & -- & -- & (5) \\
 {\sl RHESSI} & 02/02-05/08 & & \hspace*{2mm}0 & -- & -- & -- & -- & -- & \hspace*{1mm}174$^{\rm c}$ & 3.3 & 3.1 & 0.5 & \hspace*{1mm}0.3 & 0.7 & (6) \\
 {\sl RHESSI} & 02/02-05/08 & & \hspace*{1mm}97$^{\rm c}$ & 5.0 & 4.7 & 2.0 & -0.8 & \hspace*{1mm}0.3 & \hspace*{1mm}137$^{\rm c}$ & 4.3 & 3.6 & 1.8 & \hspace*{1mm}0.8 & 1.1 & (7) \\
 \hline
\end{tabular}
\begin{list}{}{}
 \item[$^{\rm a}$] N -- number of events; Mn -- mean, Md -- median, SD -- standard deviation, K --
 kurtosis; S -- skewness
 \item[$^{\rm b}$] (1) -- Roy \& Datlowe (1975); (2) -- Mariska \& McTiernan
 (1999); (3) -- Petrosian et al. (2002); (4) -- this issue; (5) -- Krucker \& Lin
 (2008); (6) -- Saint-Hilaire et al. (2008); (7) -- Ciborski \& Tomczak (2009)
 \item[$^{\rm c}$] imaging spectroscopy
 \item[$^{\rm d}$] events uncharged with the thermal component
\end{list}
\end{flushleft}
\end{table*}

Table\,\ref{tab:comp} contains two kinds of data. Papers (1), (2),
(4), and (5) have gathered a set of partially occulted flares and in
this way inform about spectra of coronal hard X-ray sources. These
papers, in exception of the last one, present also, for a
comparison, spectra for a set of non-occulted flares. They are
actually a mixture of emission of loop-top and footpoint sources but
it is believed that their spectra are dominated, especially in
higher energies, by footpoint sources. Papers (3) and (7) basing on
the imaging spectroscopy give characteristics of spectra of loop-top
and footpoint sources of the same flares. Paper (6) also used the
imaging spectroscopy but was limited to footpoint sources.

Histograms of values of the power-law index are represented in
Table\,\ref{tab:comp} by routine statistical parameters: mean,
median, standard deviation, kurtosis and skewness. In papers (1) and
(6) the source data was not published, therefore their statistical
parameters were calculated under assumption that all events in a
particular bin were equal to the middle value. The number of events
means the number of flares or the number of hard X-ray sources in
case of the imaging spectroscopy. For \yohkoh observations energy
ranges of spectra fitting are given. In case of other observations
basically the full available energy coverage were used, in exception
of paper (6) in which the power-law was fitted above the cut-off
energy.

The histograms gathered in Table\,\ref{tab:comp} show different
shapes for loop-top and footpoint sources. On the other hand, the
histograms obtained for the same group of sources show similar
shapes. Distributions of spectra of the loop-top sources show
basically broad maxima which are almost symmetrical with a small
excess for flatter spectra. Distributions of spectra of the
footpoint sources show generally narrower maxima and are strongly
asymmetric with an excess for steeper spectra.

To describe the most characteristic values of $\gamma$ we use
medians, because of the large skewness of distributions, especially
for footpoint sources. The broad range of medians within both groups
of sources is seen. This wide spread can be caused by many factors
like: properties of instruments, methods of spectra selection and
fitting, or criteria of selection of flares. Nevertheless, all
performed surveys show that hard X-ray spectra of loop-top sources
are systematically steeper than spectra of footpoint sources.

The medians obtained from low-energy bands of the HXT are especially
shifted towards the higher values, due to a strong contamination
with the thermal component. Moreover, in paper (2) very long times
of integration additionally favor an influence of the thermal
component. To avoid this contamination it is recommended to shift
the spectral fitting towards higher energies, however this activity
limits a number of considered events and charges the results with a
selection effect. The comparison of slopes in different energy
ranges of the hard X-ray spectrum, performed in Figs\,\ref{g1g2} and
\ref{g2g3}, allowed us to isolate some flares which were not charged
with the thermal component even in lower energies. Thanks to that we
have improved the statistics and our results look reasonably in
comparison with medians obtained from surveys other than {\yohkoh}.

Medians of histograms from papers (1), (5), (6), and (7) concentrate
between 4.7--5.5 and 3.1--3.8 for loop-top and footpoint sources,
respectively. However, even these values are not completely free
from some systematic charges. The spectrum of the non-occulted flare
is a mixture of photons emitted from footpoints and from the coronal
part. Despite an usual domination of footpoints, the usage of the
spectrum of the non-occulted flare as a representation of footpoint
sources introduces some bias. On the other hand, imaging
spectroscopy can distort spectra of fainter sources (usually the
loop-top ones) in the presence of brighter sources (usually the
footpoint ones). To overcome this limit partially occulted flares
are chosen, although a not precise enough selection can be
confusing. This is probably the case for paper (1) in which limb
flares ($60^{\circ}<{\theta}<90^{\circ}$) show similar values of
$\gamma$ to those obtained for partially occulted flares. In paper
(6) the distribution at high $\gamma$ is distorted by observational
bias because only flares with hard spectra (above 50\,keV) were
studied.

We have found the presence of the non-thermal component in 11 from
23 ($\sim$48\%) of partially occulted flares. In next 10 from 23
flares ($\sim$43\%), some impulsive non-thermal episodes similar to
those described by \citet{tom01} are seen, but the impulses modify
only slightly thermal appearance of these flares. In earlier surveys
the non-thermal component was discovered in 25 from 37 \citep{r+d75}
and in 50 from 55 flares \citep{k+l08} i.e. in $\sim$68\% and
$\sim$91\% of events, respectively. We estimated also the frequency
of occurrence of the non-thermal component for loop-top sources
described by \citet{pet02} comparing values of $\gamma$ in 14-34 and
23-53\,keV energy ranges. The result is $\sim$65\% (11 from 17
events).

The spatial situation of the non-thermal and thermal components is
common for the majority of investigated flares. The same rule can be
concluded from the earlier papers \citep{tom01,k+l08}. \citet{mas94}
found that the coronal hard X-ray sources situated above the soft
X-ray loop-top kernels show spectra which are as flat as the spectra
of the footpoint sources. In our survey, flares, for which the hard
X-ray and soft X-ray components are spatially separated, do not show
unusually flat spectra. The extremely shifted event in the survey of
\citet{k+l08}, the flare of 2003 November 18, also shows a modest
value of $\gamma$ equal 5.6.

What do the spectra of hard X-ray sources teach us about non-thermal
electron beams in solar flares? In the simplest scenario, the same
electron beam, described by the power-law index $\delta$, radiates
hard X-ray photons in the thin-target loop-top and thick-target
footpoint sources. Spectra emitted by these two kinds of sources
have the indices that differ by 2: ${\gamma}_{LT} = \delta + 1$ and
${\gamma}_{FP} = \delta - 1$. \citet{mar96} have obtained such a
difference and gave the interpretation for partially occulted and
non-occulted flares, but a small number of investigated events (4 +
4) needed a verification for better statistics.

Without any exception, the surveys presented in
Table\,\ref{tab:comp} confirm that the loop-top sources are
systematically steeper than the footpoint ones, i.e.
$\bar{{\gamma}}_{LT}
> \bar{{\gamma}}_{FP}$. In this paper we have obtained differences between
medians of distributions of $\gamma$ values, $\bar{{\gamma}}_{LT} -
\bar{{\gamma}}_{FP}$, to be about 3. However, the results are
seriously charged by the contamination with the thermal component.
The difference, ${\Delta}{\gamma} = \bar{{\gamma}}_{LT} -
\bar{{\gamma}}_{FP}$, for uncharged events is about 1.5. We would
like to stress that similar values of the difference
${\Delta}{\gamma}$, between 1 and 1.5, were obtained in the majority
of other surveys in Table\,\ref{tab:comp}.

The difference, ${\Delta}{\gamma} < 2$, means that a part of
investigated flares has broken the simplest scenario,
${\Delta}{\gamma} = 2$. The shape of distribution of $\gamma$
values, which is more symmetric and broader for loop-top sources in
comparison with footpoint ones, can suggest a solution. We guess
that some loop-top sources which emitted hard X-rays in the
thick-target mechanism are responsible for this. A low-value wing of
the $\gamma$ distribution for loop-top sources is occupied by flares
in which X-ray plasma ejections were observed, hence perhaps a kind
of magnetic trapping occurs. Another opportunity for the coronal
thick-target emission may be an unusually-high filling of a flare
loop with plasma \citep{kos94,v+b04}.

Values of the difference ${\Delta}{\gamma}$ higher than 2 need to
employ other explanations like, for example, separate non-thermal
electron beams responsible for loop-top and footpoint sources or
deceleration of the precipitating electrons in electric fields due
to return currents \citep{b+b08}.

A lack of correlation between the altitude and the power-law index
$\gamma$, seen in Fig.\,\ref{h-g}, suggests that the investigated
flares do not respond the overall (global) magnetic configuration of
the solar corona. For their characteristics is rather conclusive the
local magnetic configuration in which these flares were developed.

We found the non-thermal component for the majority of the
investigated flares similarly to the earlier surveys. We have
pointed out that in most cases the non-thermal and thermal
components (the hard X-ray loop-top source and the bright loop-top
kernel, respectively) are co-spatial or overlap during the impulsive
phase which confirms that both emissions come from the common plasma
volume and are strongly coupled. Events in which both components are
clearly separated, like ``Masuda flare'', occur seldom and are not
obviously characterized by a flat hard X-ray spectrum. The similar
conclusion was given by \citet{k+l08}.

We would like to stress clear inconsistency between a picture of a
flare that is shaped by statistical surveys (i.e. what typical is)
and by individual famous events (i.e. what a challenge is). On the
other hand, the models adopted to explain a specific configuration,
like in Masuda flare, are used very often for a kind of unification
i.e. for explanation of all flares. There is no doubt that frequency
of occurrence should not be a conclusive criterion to perform an
interpretation, but there is no reason to recognize such unusual
structure, like in Masuda flare, in each flare, as well.

Why Masuda-type flares are poorly represented in statistical surveys
like ours, though new examples are recently reported \citep{shi08}?
Such an unusual appearance is seen during short episodes occurring
randomly in time. On the other hand, one flare is usually
represented in catalogues by one time period around the maximum.
Thus, it is quite easy to miss such an interesting episode. For
example, the famous above-the-loop-top hard X-ray source discovered
in Masuda flare is seen at the beginning of the impulsive phase and
during the hard X-ray maximum is dominated by the strong footpoint
sources. In conclusion, searching for interesting episodes needs
careful selection of a sequence of images illustrating the
evolution.

Recently, \citet{j+t09} proposed a model of oscillating magnetic
traps in which flares appear as a cusp-like plasma volume. It
accelerates particles during the compression (hard X-ray emission),
but during the expansion is feeded with plasma coming from
chromospheric evaporation (soft X-ray emission). This model
anticipates episodes in which the magnetic mirror ratio remains
greater than 1 during the compression, so the accelerated electrons
remain there for a longer time and therefore locally generate
enhanced hard X-ray emission. Indeed, in most cases of the presented
survey we observe a clear quasi-periodicity of hard X-ray light
curves.

\begin{table*}
 \caption{List of flares showing the progressive spectral hardening}\label{tab:shh}
 \begin{center}
 \begin{tabular}{clccccc}
 \hline
 Catalogue & \hspace*{1cm}Date & GOES & Longi- & Occultation & $\gamma$ range & Time \\
 number & & class & tude & height [Mm] & & interval\\
 \hline
\hspace*{1.5mm}9700 & 1994 February 27 & M2.8 & \hspace*{-2mm}W97.5 & \hspace*{1.5mm}6.1 & 6$\rightarrow$4 & 09:09-09:18 \\
11950 & 1998 May 9 & M7.7 & W102.3 & 17.1 & 6.0$\rightarrow$3.4 & 03:29-03:39 \\
\hspace*{1mm}25540$^{\rm a}$ & 2001 April 1 & M5.5 & E107.0 & 34.1 & $>$11$\rightarrow$7.6$^{\rm b}$ & 11:48-11:54 \\
 \hline
 \end{tabular}
 \begin{list}{}{}
\item[$^{\rm a}$] Only the decay phase is available
\item[$^{\rm b}$] Strong contamination with emission of thermal
plasma
\end{list}
\end{center}
 \end{table*}

\section{Interesting groups of events} \label{inter}

In this Section short descriptions of interesting group of events
from our survey are given.

\subsection{Progressive spectral hardening}
\label{shh}

Relatively less frequent group of flares shows in hard X-rays a
progressively hardening spectral evolution that can be described
schematically by the pattern: ``soft-hard-harder'' (SHH). These
flares are very interesting due to their association with other
solar activity phenomena like: Coronal Mass Ejections, Solar
Energetic Particles, radio bursts type II and type IV. Early reports
located the radiation of the SHH flares clearly above the solar
surface, inside a postflare loop system where electrons are
accelerated and trapped \citep[][and references therein]{cli86}.
However, this picture was directly confirmed only for a few events
that occurred some distance behind the solar limb or were imaged
below 40\,keV by {\sl Hinotori}.

The \yohkoh HXT introduced a new picture of SHH flares in which
their emission is situated close to the solar surface, at flare
footpoints \citep{qiu04,tak07}. The dominance of footpoint emission
was confirmed also for some SHH flares observed by {\sl RHESSI}
\citep{sal08,g+b08}. Primarily thick-target footpoint sources emit
their hard X-rays immediately, therefore the SHH spectral pattern
cannot be explained by the trapping mechanism only. Footpoint SHH
flares strongly suggest a physical mechanism that continuously
accelerates electrons to ever higher energies.

In our survey we found three flares showing the SHH spectral
pattern. Their basic characteristics are given in
Table\,\ref{tab:shh}. All occurred far enough behind the solar limb
to eliminate the hard X-ray emission from footpoints. Thanks to
that, a better insight into the coronal part is possible.
\citet{tom08} noted that the SHH spectral pattern coincides in time
with the occurrence of a new coronal source. In case of 1998 May 9
this source  moved away gradually with the velocity increasing from
25 to 80\,km\,s$^{-1}$.

\begin{table*}
 \caption{List of \yohkoh coronal $\gamma$-ray flares}\label{gam}
 \begin{center}
 \begin{tabular}{clcccc}
 \hline
 Catalogue & \hspace*{1cm}Date & GOES & Longi- & Occultation & Photons \\
 number & & class & tude & height [Mm] & energy [keV]\\
 \hline
\hspace*{1.5mm}6240 & 1992 November 2 & X9.0 & \hspace*{-2mm}W97.9 & \hspace*{1.5mm}7.4 & \hspace*{1.5mm}$<$600$^{\rm a}$ \\
11650 & 1998 April 23 & X1.2 & E103.7 & 21.4 & $<$250 \\
11950 & 1998 May 9 & M7.7 & W102.3 & 17.1 & $<$200 \\
13650 & 1998 November 24 & X1.0 & W103.1 & 21.5 & \hspace*{1.5mm}$<$600$^{\rm a}$ \\
26060 & 2001 April 18 & C2.2 & W116.9 & 90.5 & $<$250 \\
 \hline
 \end{tabular}
 \begin{list}{}{}
\item[$^{\rm a}$] Only the decay phase is available
\end{list}
\end{center}
 \end{table*}

The progressive spectral hardening in flares from
Table\,\ref{tab:shh} was always preceded by a phase of the
``soft-hard-harder'' (SHH) spectral evolution pattern in which
another hard X-ray source was seen. Such a behavior strongly
suggests that a switch between the SHS and SHH spectral pattern
needs a serious reconfiguration of magnetic structure of a flare.

\subsection{Coronal $\gamma$-ray sources}
\label{gamma}

Thanks to RHESSI imaging capabilities it is possible to obtain hard
X-ray images in higher energies than earlier. The results limited by
counting statistics and dynamic range show that $\gamma$-rays are
emitted usually from footpoints \citep{sal08} though an example of
the coronal source is also known \citep{kru08b}. \yohkoh could not
image such high energies, however we found in the HXT Flare
Catalogue five events, recorded by the WBS, that occurred behind the
solar limb and which spectra reach at least 200\,keV (see
Table\,\ref{gam}).

All these events occurred far enough behind the solar limb to be
sure that the high-energy photons are emitted from the corona. In
this way, without imaging, we obtain an independent confirmation
that $\gamma$-rays can be produced in the corona. The flare of 1998
November 24, which produced 600-keV photons about 3 minutes after
its maximum, attracts special attention. According to the CGRO/BATSE
light curve, the flux decreased almost twice during this time
interval, hence in the maximum the photons more energetic than
600\,keV were emitted.

\citet{kru08b} interpret the coronal $\gamma$-ray emission as
relativistic electron-electron bremsstrahlung at energies perhaps of
a few~MeV. Therefore, these observations directly imply that
flare-accelerated MeV~electrons reside stably in the corona, losing
their energy collisionally and producing $\gamma$-ray continuum.

\section{Conclusions} \label{concl}

Among from 1286 flares in the \yohkoh Hard X-ray Telescope Flare
Catalogue, for which the hard X-ray images had been enclosed, we
identified 98 events that occurred behind the solar limb. Obscurity
of footpoints, that are usually brighter in hard X-rays, allowed us
to isolate the coronal parts of these flares for a more detailed
analysis. We investigated hard X-ray spectra and spatial structure
of partially occulted flares. We found that in most cases their hard
X-ray spectra consists of two co-spatial components, non-thermal and
thermal.

We note that the spectra of partially occulted flares are
systematically steeper than spectra of non-occulted events in the
Catalogue. It shows that the hard X-ray emission in solar flares is
usually less energetic in the coronal part (loop-top sources) and
more energetic close to the solar surface (footpoint sources). The
difference between medians of values of the power-law index for both
classes of sources, ${\Delta}{\gamma} = \bar{{\gamma}}_{LT} -
\bar{{\gamma}}_{FP}$, is about 3. However, this result is strongly
affected with thermal emission. For events unbiased with the thermal
component the difference ${\Delta}{\gamma}$ is equal 1.5. The
obtained value is similar to results of other surveys.

We judge that different slopes of the non-thermal component are
caused basically by different mechanisms of emission: thin-target
for loop-top sources and thick-target for footpoint sources. The
difference of ${\Delta}{\gamma}$ smaller than 2 suggests some
exceptions from this rule. We have presented some arguments that a
part of coronal hard X-ray sources is thick-target instead of to be
thin-target, e.g. several unusually high-energetic events which has
occurred in configuration suggesting a magnetic trapping. An
additional argument, that for characteristics of flares conclusive
is local magnetic configuration in which they develops, is a lack of
correlation between the altitude of flares and the hard X-ray
power-law index $\gamma$.

Further good observations and precise analysis are needed to
arbitrate about details of non-thermal electrons acceleration and
propagation in solar flares. Detailed imaging spectroscopy of
individual events as well as massive surveys of many flares are
desired.

\begin{acknowledgements}

The \yohkoh satellite is a project of the Institute of Space and
Astronautical Science of Japan. I am grateful to late Professor
Takeo Kosugi, Dr. Jun Sato, and their collaborators for preparing
the \yohkoh HXT Flare Catalogue which made this work easier. I thank
also my colleagues from a working group of the International Space
Science Institute (Bern, Switzerland), dedicated to coronal hard
X-ray sources, for a fruitful discussion. This work was supported by
Polish Ministry of Science and High Education grant No.
N\,N203\,1937\,33.

\end{acknowledgements}

\onecolumn \setlongtables
\begin{small}
\begin{longtable}{rclccccrrrrrrl}
\caption{List of partially occulted flares found in the YOHKOH HXT
Flare Catalogue$^{\rm a}$.} \label{occlist}
\\
\hline
(1) & (2) & (3) & (4) & (5) & (6) & (7) & (8) & (9) & (10) & (11) & (12) & (13) & (14) \\
\hline
\endfirsthead
\caption{cont.}\\
\hline
(1) & (2) & (3) & (4) & (5) & (6) & (7) & (8) & (9) & (10) & (11) & (12) & (13) & (14) \\
 \hline
 \endhead
00180 & 1991/10/21 & 12:51:00+38 & C7.8 & 6891 &
S12E92.5 & 0.8 & 4.6 & 1.0 & 2.5 & 8.0 & ... & ... & e,o,s \\
01050 & 1991/11/30 & 19:07:01+288 & M1.0 & 6952 &
N21E108.4 & 1.2 & 34.4 & 1.0 & 42.0 & 9.4 & ... & ... & m \\
01060 & 1991/12/02 & 04:53:11+76 & M3.6 & 6952 & N18E91.5
& 2.4 & 2.8 & 4.3 & 7.9 & 5.5 & 5.0 & ... & m \\
01320 & 1991/12/09 & 02:03:08+57 & M1.4 & 6966 &
S06E91.0 & 1.0 & 1.9 & 2.0 & 3.7 & 9.5 & ... & ... & e,o,s \\
01340 & 1991/12/09 & 04:19:09+242.5 & M3.6 &
6966 & S06E89.8 & 1.0 & 0 & 3.5 & 6.2 & 10.6 & ... & ... & e,m,o \\
02270 & 1992/01/13 & 19:04:35+349.5 & M1.3 & 7012 &
S10E95.3 & 0.9 & 10.1 & 4.9 & 11.8 & 9.9 & ... & ... & m \\
02860 & 1992/02/06 & 20:51:22+240.5 & M4.1 & 7030 &
N05W93.8 & 3.0 & 6.8 & 4.6 & 9.7 & 8.7 & ... & ... & m \\
03270 & 1992/02/19 & 14:45:14+429 & C9.4 & 7067 & N06E94.4
& 1.8 & 8.4 & 3.0 & 7.4 & 8.5 & ... & ... & m \\
04270 & 1992/06/28 & 04:47:05+153.5 & X1.8 &
7205 & N12W101.5 & 2.2 & 21.3 & 14.8 & 41.4 & 6.3 & 6.4 & ... & e,s \\
05090 & 1992/08/25 & 19:02:52+160 & C8.7 & 7260 & N13W98.2
& 2.0 & 15.2 & 14.1 & 32.7 & 8.5 & ... & ... & s \\
05210 & 1992/09/06 & 09:20:11+27 & C8.3 & 7276 & N15E101.4
& 0.5 & 21.1 & 1.0 & 16.3 & 6.1 & ... & ... & s \\
05260 & 1992/09/06 & 23:37:17+29 & M1.3 & 7276 &
N17E93.6 & 2.6 & 6.4 & 1.0 & 3.2 & 8.3 & 7.5 & ... & o,s \\
05590 & 1992/10/05 & 09:24:28+47.5 & M2.0 & 7293 &
S08W89.7 & 3.9 & 0 & 2.0 & 3.6 & 4.3 & 4.3 & ... & o,s \\
05920 & 1992/10/27 & 22:17:44+104.5 & C5.4 & 7315 &
N07W88.5 & 1.5 & 0 & 4.1 & 7.5 & 7.6 & ... & ... & m,o \\
06230 & 1992/11/01 & 11:44:06+53.5 & C4.9 & 7321 &
S25W89.3 & 2.2 & 0 & 11.9 & 21.2 & 5.9 & 8.7 & ... & o,s \\
06240 & 1992/11/02 & 02:59:49+42 &
X9.0 & 7321 & S25W97.9 & 2.3 & 14.2 & 14.6 & 33.6 & 8.0 & 7.3 & 6.0 & e,g,s \\
06280 & 1992/11/05 & 20:30:17+167.5 & C8.2 & 7323 &
S17W91.5 & 2.0 & 2.6 & 5.0 & 9.1 & 8.7 & ... & ... & s \\
06440 & 1992/11/24 & 10:01:02+157 & C6.9 & 7342 & S07W95.5
& 5.0 & 7.4 & 3.0 & 8.6 & 6.5 & ... & ... & s \\
06480 & 1992/11/24 & 20:31:03+271.5 & C6.4 &
7342 & S07W101.2 & 5.2 & 17.9 & 1.5 & 16.3 & 8.7 & ... & ... & e,m \\
06810 & 1993/02/01 & 01:57:59+151 & M2.2 & 7416 &
S10E92.6 & 1.7 & 4.6 & 1.0 & 2.5 & 6.9 & 4.6 & ... & m,o \\
06820 & 1993/02/01 & 06:58:25+119.5 & M1.4 &
7416 & S09E89.8 & 2.3 & 0.0 & 0.5 & 0.9 & 9.9 & ... & ... & e,m,o \\
07580 & 1993/03/15 & 19:07:00+138.5 & C5.5 & 7440 &
S06W91.6 & 2.0 & 3.0 & 3.5 & 6.5 & 7.5 & ... & ... & m \\
07590 & 1993/03/15 & 21:05:53+479 & M3.0 & 7440
& S03W92.7 & 2.0 & 4.9 & 29.5 & 53.3 & 10.5 & ... & ... & e,m \\
07680 & 1993/03/24 & 03:21:45+42 & C6.6 & 7448 & N15W92.1
& 3.7 & 3.8 & 2.7 & 5.3 & 6.2 & 7.4 & ... & s \\
08220 & 1993/06/24 & 14:54:48+176 & C5.9 & 7530 &
S11E90.9 & 0.7 & 1.7 & 1.0 & 1.9 & 9.8 & ... & ... & m,o \\
08230 & 1993/06/24 & 17:27:32+132.5 & M4.2 &
7530 & S11E89.5 & 0.7 & 0 & 4.6 & 8.2 & 10.8 & ... & ... & e,s \\
08460 & 1993/09/26 & 10:26:04+63 & C3.4 & 7590 &
N14E98.0 & 1.1 & 14.0 & 1.0 & 8.8 & 7.1 & ... & ... & e,s \\
08470 & 1993/09/26 & 17:26:03+61 & C3.4 & 7590 & N14E94.0
& 1.1 & 7.0 & 1.0 & 3.5 & 7.6 & ... & ... & s \\
08480 & 1993/09/26 & 18:29:18+54.5 & C2.6 & 7590 &
N14E93.4 & 1.1 & 5.9 & 1.1 & 3.2 & 5.6 & 5.7 & ... & s \\
09650 & 1994/01/29 & 04:10:16+132 & M2.4 & 7654
& N05W92.9 & 1.9 & 5.2 & 3.0 & 6.2 & ... & ... & ... & c,s \\
09660 & 1994/01/29 & 11:24:49+80.5 & M2.4 & 7654 &
N07W97.0 & 1.9 & 12.4 & 6.5 & 16.9 & 7.0 & 8.1 & ... & m \\
09700 & 1994/02/27 & 09:06:47+541 & M2.8 & 7671
& N08W97.5 & 3.8 & 13.6 & 12.6 & 38.6 & 7.9 & ... & ... & h,m \\
11090 & 1997/11/25 & 05:34:45+162.5 & C5.0 & 8113 &
N21E96.4 & 0.7 & 12.2 & 8.9 & 20.6 & 8.0 & ... & ... & s \\
11100 & 1997/11/25 & 14:40:44+37.5 & C1.1 & 8113 &
N21E91.6 & 0.6 & 3.1 & 0.5 & 1.2 & 5.6 & ... & ... & o,s \\
11290 & 1998/01/03 & 17:13:25+127 & M2.7 & 8124 &
S20W102.0 & 2.8 & 22.0 & 3.5 & 22.9 & 7.1 & 7.8 & ... & m \\
11510 & 1998/03/23 & 02:59:31+121.5 & M2.3 & 8179 &
S22W99.3 & 1.9 & 16.6 & 15.7 & 38.3 & 8.4 & ... & ... & s \\
11530 & 1998/03/24 & 01:47:46+60 & C2.3 & 8180 & S32W98.9
& 3.0 & 16.7 & 10.8 & 29.4 & 5.8 & ... & ... & s \\
11540 & 1998/03/24 & 04:42:15+75 & C4.3 & 8180 & S32W100.5
& 3.0 & 19.7 & 10.0 & 32.0 & 6.3 & ... & ... & s \\
11650 & 1998/04/23 & 05:39:55+151 & X1.2 & 8210
& S18E103.7 & 0.7 & 25.5 & 2.2 & 30.0 & 7.7 & 7.7 & 2.4 & g,s \\
11660 & 1998/04/24 & 08:46:23+74.5 & C8.9 & 8210 &
S20E91.3 & 2.2 & 2.3 & 13.5 & 24.2 & 4.7 & 5.1 & ... & s \\
11910 & 1998/05/08 & 14:20:54+67.5 & M1.8 & 8210
& S17W95.2 & 1.5 & 9.7 & 7.4 & 16.2 & 10.2 & ... & ... & e,s \\
11930 & 1998/05/09 & 00:17:53+22.5 & C8.3 & 8210 &
S14W100.6 & 1.6 & 19.7 & 1.1 & 14.4 & 6.3 & ... & ... & s \\
11950 & 1998/05/09 & 03:22:56+70.5 & M7.7 &
8210 & S17W102.3 & 1.6 & 22.8 & 1.4 & 19.6 & 6.8 & 6.1 & 3.9 & g,h,s \\
12000 & 1998/05/10 & 08:26:33+113 & M1.6 & 8220
& S27E91.4 & 2.4 & 2.6 & 4.5 & 8.2 & 9.8 & ... & ... & e,s \\
12010 & 1998/05/10 & 13:18:30+14 & M3.9 & 8220 &
S27E88.8 & 2.4 & 0 & 1.1 & 2.1 & 4.9 & 3.5 & 3.3 & e,o,s \\
13580 & 1998/11/22 & 12:21:38+317.5 & C8.8 &
8393 & S17E93.4 & 4.2 & 6.2 & 3.0 & 6.6 & 8.8 & ... & ... & m \\
13610 & 1998/11/23 & 05:58:41+39.5 & C4.9 & 8384 &
S29W92.3 & 2.0 & 4.3 & 4.2 & 8.1 & 6.1 & 7.5 & ... & s \\
13620 & 1998/11/23 & 06:53:06+5.5 & X2.2 & 8384
& S29W92.8 & 2.0 & 5.2 & 8.3 & 15.7 & 10.3 & ... & ... & e,s \\
13650 & 1998/11/24 & 02:17:48+316.5 &
X1.0 & 8384 & S30W103.1 & 2.0 & 24.5 & 3.2 & 27.3 & 8.8 & 9.6 & 4.5 & e,g,m \\
15640 & 1999/06/17 & 17:20:47+56.5 & M3.6 & 8584
& N22W?? & ... & $>$0 & 5.9 & $>$10.5 & 7.7 & ... & ... & e,s \\
16290 & 1999/07/23 & 05:01:27+44 & C9.4 & 8645 & S23E97.4
& 3.3 & 13.8 & 4.0 & 13.5 & 6.8 & ... & ... & s \\
16310 & 1999/07/23 & 15:56:13+21.5 & M1.0 & 8645 &
S26E91.6 & 3.3 & 3.0 & 1.1 & 2.3 & 5.5 & 6.5 & ... & o,s \\
16810 & 1999/08/07 & 19:13:10+309 & M1.2 & 8645
& S20W103.1 & 3.3 & 24.4 & 9.3 & 36.8 & 10.9 & ... & ... & e,s \\
16820 & 1999/08/07 & 20:54:08+741 & M1.7 & 8645 &
S29W104.0 & 3.3 & 26.1 & 3.9 & 31.5 & 10.0 & ... & ... & m \\
17400 & 1999/10/01 & 00:15:28+60.5 & C7.7 & 8716 &
N22E95.8 & 2.2 & 10.7 & 1.1 & 5.8 & 7.8 & ... & ... & s \\
17710 & 1999/10/27 & 09:09:39+90.5 & M1.0 & 8737 &
S12W87.8 & 5.9 & 0 & 1.8 & 3.2 & 6.0 & 4.4 & ... & o,s \\
17720 & 1999/10/27 & 13:28:21+535 & M1.8 & 8737 & S15W90.3
& 6.0 & 0.6 & 4.0 & 7.1 & 9.1 & ... & ... & m \\
17730 & 1999/10/27 & 15:28:00+762 & M1.4 & 8737 & S14W91.6
& 6.0 & 2.6 & 11.5 & 20.7 & 9.9 & ... & ... & m \\
17790 & 1999/11/05 & 18:16:47+34 & M3.0 & 8759 &
N12E95.8 & 1.3 & 11.0 & 7.0 & 16.2 & 8.4 & ... & ... & e,s \\
18610 & 1999/12/17 & 00:19:22+74 & C7.0 & 8806 &
N19E91.8 & 0.7 & 3.4 & 1.0 & 2.1 & 7.9 & 7.8 & ... & o,s \\
19030 & 2000/01/18 & 09:37:03+299 & M1.2 & 8827 &
S15W105.6 & 3.1 & 28.2 & 1.8 & 30.9 & 6.8 & 7.0 & ... & m \\
19630 & 2000/03/07 & 03:48:58+24 & C2.9 & 8906 &
S16E95.9 & 6.7 & 10.7 & 1.6 & 6.7 & 7.0 & ... & ... & o,s \\
20920 & 2000/05/03 & 22:58:52+741 & M1.1 & 8970 & S18W90.1
& 2.2 & 0.0 & 7.4 & 13.2 & 9.3 & ... & ... & m \\
20930 & 2000/05/05 & 15:18:42+1954 & M1.5 & 8970 &
S18W110.1 & 3.7 & 37.5 & 2.2 & 51.6 & 8.5 & 4.7 & ... & m \\
20980 & 2000/05/12 & 08:41:48+85 & C8.1 & 8998 & S14E90.5
& 3.9 & 0.9 & 3.6 & 6.4 & 9.1 & ... & ... & s \\
21030 & 2000/05/13 & 01:36:19+232 & M1.1 & 9002
& N22E108.8 & 0.8 & 33.8 & 4.9 & 51.5 & 9.5 & ... & ... & e,s \\
21050 & 2000/05/14 & 00:25:08+97.5 & C7.5 & 9002
& N22E96.1 & 0.7 & 11.0 & 4.5 & 12.3 & 7.2 & ... & ... & e,s \\
21280 & 2000/05/17 & 04:02:46+14 & C7.1 & 8993 &
S20W90.4 & 1.7 & 0.7 & 2.0 & 3.6 & 4.6 & 5.0 & ... & o,s \\
21530 & 2000/05/24 & 03:13:26+83.5 & C7.0 & 9017 &
S12E92.9 & 3.0 & 5.3 & 2.1 & 4.6 & 7.9 & ... & ... & o,s \\
21880 & 2000/06/12 & 01:37:32+10 & C6.1 & 9042 &
N21E91.2 & 7.8 & 2.2 & 1.6 & 3.0 & 5.8 & 4.1 & ... & o,s \\
22570 & 2000/07/15 & 04:33:29+32 & C2.7 & 9090 &
N14E94.8 & 0.8 & 8.5 & 0.5 & 3.4 & ... & ... & ... & l,s \\
23070 & 2000/07/27 & 16:46:58+111 & M1.5 & 9087 &
S09W105.4 & 1.2 & 26.9 & 5.7 & 36.7 & 7.2 & 8.4 & ... & s \\
23590 & 2000/09/22 & 23:46:52+69 & C8.5 & 9165 & N14W94.5
& 1.5 & 8.0 & 4.6 & 10.4 & 5.6 & 6.6 & ... & s \\
23750 & 2000/09/30 & 20:15:37+17.5 & M1.8 & 9169
& N07W90.8 & 2.3 & 1.4 & 3.6 & 6.5 & 9.2 & ... & ... & e,s \\
23780 & 2000/10/01 & 07:01:39+26 & M5.0 & 9169 &
N08W96.8 & 2.4 & 12.2 & 10.6 & 23.9 & 10.2 & ... & ... & e,s \\
23790 & 2000/10/01 & 14:00:08+56 & M2.2 & 9169 & N09W100.7
& 2.4 & 19.2 & 2.1 & 16.3 & 6.0 & 6.6 & ... & m \\
23810 & 2000/10/02 & 02:17:44+23 & C5.0 & 9182 &
N01E89.6 & 1.4 & 0 & 2.0 & 3.6 & 4.8 & 4.6 & ... & o,s \\
23940 & 2000/10/16 & 05:41:56+166 & C7.0 & 9182 &
N04W107.4 & 3.3 & 30.1 & 3.0 & 39.0 & 7.1 & 9.6 & ... & m \\
23950 & 2000/10/16 & 06:59:48+1018.5 & M2.5 & 9182 &
N03W108.3 & 3.3 & 31.5 & 4.0 & 44.6 & 10.8 & ... & ... & s \\
24480 & 2000/12/08 & 16:18:57+90.5 & C4.3 & 9246 &
S09W90.6 & 1.3 & 1.1 & 2.1 & 3.8 & 7.0 & ... & ... & m,o \\
24580 & 2000/12/18 & 20:12:57+27 & C2.7 & 9280 &
N08E92.5 & 0.5 & 4.4 & 2.0 & 4.2 & 9.3 & ... & ... & o,s \\
24840 & 2001/01/03 & 18:05:31+44.5 & C2.7 & 9302
& N20E97.2 & 1.2 & 12.8 & 9.0 & 22.0 & ... & ... & ... & c,s \\
25540 & 2001/04/01 & 11:42:36+36 &
M5.5 & 9415 & S21E107.0 & 0.7 & 30.3 & 23.6 & 77.9 & 9.2 & 6.7 & ... & e,h,s \\
25870 & 2001/04/05 & 02:06:55+75 & M3.1 & 9393 & N15W99.2
& 1.6 & 16.3 & 3.2 & 15.2 & 6.1 & 6.8 & ... & s \\
25880 & 2001/04/05 & 05:12:45+119 & M1.1 & 9393 &
N15W100.8 & 1.6 & 19.3 & 4.6 & 21.3 & 8.1 & ... & ... & m \\
25890 & 2001/04/05 & 08:33:08+8 & M8.4 & 9393 &
N14W102.9 & 1.7 & 23.1 & 6.3 & 30.1 & 8.3 & 6.6 & ... & e,s \\
26060 & 2001/04/18 & 02:14:29+36.5 & C2.2 & 9415
& S21W116.9 & 1.8 & 47.9 & 9.0 & 108.4 & 3.4 & 3.7 & 3.7 & g,s \\
26480 & 2001/05/20 & 06:02:17+6.5 & M6.4 & 9455 &
S17W94.4 & 2.6 & 8.1 & 0.7 & 3.4 & 4.0 & 3.4 & 3.6 & o,s \\
26490 & 2001/05/20 & 09:19:43+51.5 & M1.5 & 9455 &
S20W96.2 & 2.6 & 11.3 & 1.0 & 6.1 & 8.5 & 5.8 & ... & m \\
27080 & 2001/08/09 & 00:39:15+120 & C6.0 & 9557 & S19W91.7
& 3.3 & 3.1 & 5.4 & 9.9 & 6.6 & 7.9 & ... & s \\
27400 & 2001/08/29 & 09:12:02+59.5 & C6.5 & 9587 &
S12W97.4 & 3.5 & 12.7 & 1.0 & 7.8 & 6.7 & 8.3 & ... & s \\
27500 & 2001/09/03 & 01:54:38+133 & C9.0 & 9607 & S15E93.1
& 3.3 & 5.6 & 3.4 & 7.1 & 10.1 & ... & ... & s \\
27540 & 2001/09/03 & 18:23:03+193.5 & M2.5 & 9608 &
S22E96.2 & 1.0 & 12.2 & 1.4 & 6.9 & 8.1 & 6.9 & ... & m \\
27890 & 2001/09/11 & 07:00:00+69 & M1.2 & 9616 & S10E92.3
& 4.4 & 4.1 & 6.8 & 12.7 & 8.0 & ... & ... & s \\
28260 & 2001/09/19 & 06:54:11+56.5 & C7.2 & 9608 &
S23W93.9 & 2.2 & 7.6 & 10.2 & 19.9 & 5.6 & 7.2 & ... & s \\
29110 & 2001/10/29 & 01:56:00+41.5 & M1.3 & 9669 &
N12W103.9 & 3.0 & 25.5 & 2.5 & 26.0 & 5.3 & 5.0 & ... & m \\
30430 & 2001/12/01 & 15:12:21+847 & M4.8 & 9727
& S25E97.8 & 2.4 & 14.3 & 13.6 & 31.6 & 9.0 & ... & ... & e,m \\
30470 & 2001/12/02 & 21:29:50+251.5 & M2.0 & 9714 &
S11W108.1 & 2.9 & 32.5 & 8.6 & 53.0 & 9.7 & ... & ... & m \\
\hline
\end{longtable}
\begin{list}{}{}
\item[$^{\rm a}$] Descriptions: (1) - catalogue event number; (2) - date
(YYYY/MM/DD); (3) - time integration (start time [UT] + duration in
seconds); (4) - GOES X-ray class; (5) - NOAA active region number;
(6) - location in solar coordinates; (7) - uncertainty of longitude
estimation; (8) - estimated number of hours before/after the limb
passage; (9) - altitude of HXR centroid above the solar limb in SXT
pixels [2.45 arcsec]; (10) - estimated total height in 10$^3$ km;
(11) - ${\gamma}1$, the power-law index from hardness ratio M1/L;
(12) - ${\gamma}2$, the power-law index from hardness ratio M2/M1;
(13) - ${\gamma}3$, the  power-law index from hardness ratio H/M2;
(14) - remarks: c - corrupted HXT file (no HXT diagnostics), e -
maximum of flare outside the time integration, g - gamma rays, h -
progressive spectral hardening, l - only channel L (no HXT
diagnostics), m - multiple bursts, o - event situated too close to
the solar limb, possible emission overshooting from footpoints, s -
single burst.
\end{list}
\end{small}
\twocolumn

\end{document}